# Cosmological Problem Solution by Complex-Dynamical Interaction Analysis[*]

## A. P. Kirilyuk

*G. V. Kurdyumov Institute for Metal Physics,*
*National Academy of Sciences of Ukraine,*
*36 Academician Vernadsky Blvd.,*
*UA-03142 Kyyiv, Ukraine*

Universe structure emerges in the unreduced, complex-dynamical interaction process with the simplest initial configuration (two attracting homogeneous fields). The unreduced interaction analysis avoiding any perturbative model gives intrinsically creative cosmology describing the real, explicitly emerging world structure with dynamic randomness at all levels. Without imposing any postulates or additional entities, we obtain physically real, three-dimensional space, irreversibly flowing time, elementary particles with their detailed structure and intrinsic properties, causally complete and unified version of quantum and relativistic behaviour, the origin and number of naturally unified fundamental forces, classical behaviour emergence in a closed system, and true quantum chaos. Major problems of standard cosmology and astrophysics are consistently solved in this extended picture, including those of quantum cosmology and gravity, entropy growth and time arrow, "hierarchy" of elementary particles (Planckian unit values), "anthropic" difficulties, Big Bang contradictions, and "missing" ("dark") mass and energy. Universality of the proposed theory is explicitly expressed by the symmetry (conservation and transformation) of dynamic complexity providing the unified, irregularly structured, but always exact (never "broken") Order of the World that underlies all universe structures, phenomena and laws.

Структура Всесвіту виникає в процесі нередукованої, складно-динамічної взаємодії з найпростішою початковою конфігурацією (два однорідних поля що взаємно

---

притягаються). Повний аналіз нередукованої взаємодії, який уникає будь-яких пертурбативних моделей, призводить до внутрішньо творчої космології, яка описує реальну, безпосередньо виникаючу структуру світу с динамічною випадковістю на всіх рівнях. Без нав'язування будь-яких постулатів та додаткових сутностей, ми одержуємо фізично реальний тримірний простір, необоротно спливаючий час, елементарні частки з їх детальною структурою та внутрішніми властивостями, каузально повну та об'єднану версію квантової і релятівістської поведінки, походження та число природно об'єднаних фундаментальних взаємодій, виникнення класичної поведінки в замкненій системі та істинний квантовий хаос. У цій самоузгодженій картині послідовно вирішені основні проблеми стандартної космології і астрофізики, включаючи трудності квантової космології та гравітації, зростання ентропії та стрілу часу, «ієрархію» елементарних часток (значення планковських одиниць), «антропні» проблеми, протиріччя Великого вибуху та «недостатню» («темну») масу і енергію. Універсальність запропонованої теорії безпосередньо виражається симетрією (збереженням та перетворенням) динамічної складності, яка дає єдиний, нерегулярно структурований, але завжди точний (ніде не «порушений») Світовий Порядок що лежить в основі усіх структур, явищ і законів Всесвіту.

Структура Вселенной возникает в процессе нередуцированного, сложно-динамического взаимодействия с простейшей начальной конфигурацией (два взаимно притягивающихся однородных поля). Полный анализ нередуцированного взаимодействия, избегающий каких-либо пертурбативных моделей, приводит к внутренне созидательной космологии, описывающей реальную, непосредственно возникающую структуру мира с динамической случайностью на всех уровнях. Без навязывания каких-либо постулатов или дополнительных сущностей, мы получаем физически реальное, трёхмерное пространство, необратимо текущее время, элементарные частицы с их детальной структурой и внутренними свойствами, каузально полную и объединённую версию квантового и релятивистского поведения, происхождение и число естественно объединённых фундаментальных взаимодействий, возникновение классического поведения в замкнутой системе и истинный квантовый хаос. В этой самосогласованной картине последовательно разрешены основные проблемы стандартной космологии и астрофизики, включая трудности квантовой космологии и гравитации, рост энтропии и стрелу времени, «иерархию» элементарных частиц (значения планковских единиц), «антропные» проблемы, противоречия Большого взрыва и «недостающую» («тёмную») массу и энергию. Универсальность предлагаемой теории непосредственно выражается симметрией (сохранением и преобразованием) динамической сложности, дающей единый, нерегулярно структурированный, но всегда точный (нигде не «нарушенный») Мировой Порядок, лежащий в основе всех структур, явлений и законов Вселенной.

# 1. COMPLEX-DYNAMICAL VS STANDARD COSMOLOGY

Contrary to experimental, observational successes in modern astrophysics, the explanatory power of respective cosmological theories remains limited, so that the number of unsolved problems only grows, while those considered to be "solved" often resemble rather a "plausibly" looking adjustment of artificially introduced, abstract entities and free parameters (see e.g. [1]). Without entering into detailed discussion of those difficulties, we only note here a possible general origin of such situation, which is inherent in the general scholar science approach, but has particularly strong manifestations in cosmology. As it was first emphasized by Bergson [2] and confirmed by further science development (see e.g. [3]), conventional science methods do *not* describe explicit structure emergence as such, but are limited instead to postulation of already existing structure configuration, properties, and simplified, imitative "evolution" (in the form of empirically guessed "laws", "principles", "models", etc.). Whereas such description can be useful in the study of simple, easily measurable and "smoothly" evolving objects (the canonical case of "Newtonian mechanics"), it should be much less efficient in explanation of the origin and dynamics of systems, such as the universe and its many-body objects, that cannot be simply "postulated" with all their observed properties because they undergo strong, *qualitative* changes of configuration (explicit emergence of structure) involving many diverse, hierarchically organised and entangled elements.

In other words, the *true* cosmology should be able to describe the unreduced, *explicit* formation of a complicated structure, which just remains obscure in the usual theory framework. A related difficulty of the latter is that it *does not* consistently *solve* any realistic, many-body *interaction problem*, always resorting to one or another simplified "model" or "perturbative" approximation, whereas it is just the *unreduced*, "nonintegrable" interaction process that underlies any real structure formation. In particular, standard theory cannot provide the unambiguous, universal origin of the major property of *mass* (and *energy*), operating instead with its measurable inertial and gravitational manifestations. Although this problem could remain among "less practically important" ones in "Newtonian" science, the difficulties with strangely "invisible", "dark" mass and energy have "suddenly" emerged now on the global scale as quite important, if not fatal, defects of the *entire* conventional world picture.

In this report we describe a *new, qualitatively extended cosmology framework* based on the *unreduced*, truly "exact" *solution of arbitrary interaction problem* that gives *explicit* emergence of *real* world structures and properties, without any artificial simplification and leads to the rigorously derived, truly *universal* concept of *dynamic complexity* [4-17]. This unreduced dynamic complexity is different from the existing mechanistic imitations of "complexity" in conventional theory and *unifies qualitatively extended* versions of dynamical chaos, self-organisation, self-organised criticality, "synchronisation", "chaos control", fractality, adaptability, etc.

We start with showing how *all* the fundamental universe entities and properties, including physically specified space and time, elementary particles, their properties, interactions and dynamics, *explicitly emerge* in the *provably simplest* initial configuration of *interaction process*, comprising two structureless, omnipresent, physically real fields, homogeneously attracted to each other (section 2). It becomes possible due to the *unreduced*, non-simplified solution of (arbitrary) interaction problem within the generalised effective potential method, giving rise to the qualitatively new, *dynamically multivalued* system configuration, consisting from its many *equally probable*, but *incompatible* versions, or *realisations*.

It is important that we obtain *together* the *main entities* (space, time, particles), their *properties* (space structure and number of dimensions, irreversible time flow, mass-energy, charge, spin, interactions), and *dynamical laws* (quantum and relativistic behaviour) within the *same*, *intrinsically unified* concept of (interaction) complexity, using a rigorous derivation procedure and *no* additional, postulated laws or entities besides the evidently "minimal" starting interaction configuration (section 3). We show then how the naturally emerging, truly *dynamic* properties of *complexity* and *chaoticity* give rise to all higher-level structures and solve the difficulties of conventional theory that neglects those major features because of its artificial reduction and *therefore* loses the essence of such basic properties as mass and energy (sections 4, 5).

We emphasize the *intrinsically unified* and *reality-based* character of the proposed solution to all major problems of usual theory, consistently derived simply due to the *unreduced, universally nonperturbative analysis* of an *arbitrary (generic) interaction problem*, which *confirms* the power of genuine, *unreduced* science and reveals the *exact origin* of the standard theory limitations and difficulties as its dynamically single-valued, zero-complexity approximation neglecting all really emerging system realisations except a single, "averaged" one. The ultimate, mathemati-cally exact expression of the obtained unification is provided by the *universal symmetry, or conservation, of complexity*, which determines the emergence and dynamics of all universe structures and therefore constitutes the genuine, unique Order of the World (section 2) [4-6].

## 2. UNIVERSE STRUCTURE EMERGENCE AS A RESULT OF UNREDUCED INTERACTION PROCESS

No structure can emerge without interaction. Consistent universe structure formation should start from the *simplest* possible (least structured) interaction configuration, which is still able to produce explicitly the observed real structures. The most structureless configuration of a physically real system with interaction is given by two homogeneous (*effectively structureless*), uniformly interacting entities represented by two *physically real* fields/media, or *protofields*, which are attracted to each other and whose detailed composition (of sufficiently small elements) does not play the key role in the following structure formation [4-6,11-17]. Efficient, structure-forming interaction between protofields supposes their different physical qualities designated as *gravitational* protofield (or medium) and *electromagnetic* (e/m) protofield, since we show later that they are responsible for the *emerging* (and universally present) gravitational and e/m interactions, respectively.

The physically real protofields are omnipresent and therefore *cannot* be related to any postulated (let alone "hidden" and abstract) spatial "dimensions", time "variables", other *mathematical* "structures", laws, etc., none of which may have a sense at this initial stage (cf. recent imitations within so-called "brane-world" scenarios of the unitary theory [18-20]). Extended, complex-dynamical and physically real versions of those entities and laws are *consistently derived* in our theory starting from the *existence equation*, which suitably generalises major dynamic equations and describes the above simplest protofield interaction without any limitation or model assumption [4-6,11-17]:

$$\left[ h_{\mathrm{g}}\left(\xi\right)+V_{\mathrm{eg}}\left(\xi,q\right)+h_{\mathrm{e}}\left(q\right)\right]\Psi\left(\xi,q\right)=E\Psi\left(\xi,q\right) , \qquad (1)$$

where $h_{\mathrm{g}}(\xi)$ and $h_{\mathrm{e}}(q)$ are "generalised Hamiltonians", representing the internal dynamical properties of the free (non-interacting) gravitational and e/m protofields in terms of a measure of the unreduced dynamic complexity defined below, $V_{\mathrm{eg}}(\xi,q)$ is the corresponding expression of (gen-erally arbitrary) potential of attractive interaction between protofields, whose physically different degrees of freedom are represented by $\xi$ (gravitational medium) and $q$ (e/m protofield), $\Psi(\xi, q)$ is the compound system (universe) *state-function* characterising completely its configuration and properties, and $E$ is the eigenvalue of the generalised Hamiltonian for the whole system. Note that eq. (1) generalising e.g. the quantum-mechanical Schrödinger equation or the classical Hamilton-Jacobi equation does not assume anything beyond the initial system configuration and can eventually take the form of various, including "nonlinear", "model" equations (although we show below, in a self-consistent way, that its "Hamiltonian" form is indeed absolutely universal [4-6,11-17]).

It is convenient to express the problem in terms of e/m protofield excitations (local deformations):

$$\Psi(\xi, q) = \sum_n \psi_n(\xi)\varphi_n(q), \quad h_e(q)\varphi_n(q) = \varepsilon_n\varphi_n(q), \qquad (2)$$

where $\{\varphi_n(q), \varepsilon_n\}$ is the complete set of orthonormal eigen-solutions for the free e/m protofield Hamiltonian $h_e(q)$. Substituting the first eq. (2) into eq. (1) and using the standard procedure of scalar-product separation (e.g. by integration), we obtain the system of equations for $\{\psi_n(\xi)\}$:

$$\left[h_g(\xi) + V_{nn}(\xi)\right]\psi_n(\xi) + \sum_{n' \neq n} V_{nn'}(\xi)\psi_{n'}(\xi) = \eta_n\psi_n(\xi), \qquad (3)$$

where $\eta_n \equiv E - \varepsilon_n$ and

$$V_{nn'}(\xi) = \int_{\Omega_q} dq\varphi_n^*(q)V_{eg}(\xi, q)\varphi_{n'}(q) . \qquad (4)$$

Note that eqs. (3) express the same problem configuration as eq. (1), but now in terms of the "physically specified" degrees of freedom of e/m protofield, which should be possible for any correct model of the protofield dynamics.

Usual, perturbative analysis of system (3) would reduce it to separated, "integrable" equations of the form

$$\left[h_g(\xi) + V_{nn}(\xi) + \tilde{V}_n(\xi)\right]\psi_n(\xi) = \eta_n\psi_n(\xi), \qquad (5)$$

where an integrable "mean-field" potential $\tilde{V}_n(\xi)$ can vary between zero and an extreme configuration, such as

$$\tilde{V}_n(\xi) = \sum_{n' \neq n} V_{nn'}(\xi) . \qquad (6)$$

If, however, we avoid any perturbative reduction of system (3) and try to find its unreduced solution by the method of substitution using the standard Green function technique, we arrive at the problem formulation in terms of generalised *optical, or effective, potential (EP)* [4-17,21,22]. The latter is a well-known method, but used in its reduced, perturbative version (see e.g. [21]). Direct analysis of the unreduced EP expression shows that the original problem has the *redundant* number of locally "complete" and thus *mutually incompatible*, but *equally real* solutions called system and problem *realisations* [4-17,22]. Therefore the truly *complete general solution* to a problem is given, in terms of system "density" $\rho(\xi,q)$ (generalising all measured quantities), by the *causally probabilistic* sum over redundant realisations, which *permanently replace one another* in a *dynamically random* order thus defined:

$$\rho(\xi,q) = \left|\Psi(\xi,q)\right|^2 = \sum_{r=1}^{N_\Re \oplus} \rho_r(\xi,q), \quad \rho_r(\xi,q) = \left|\Psi_r(\xi,q)\right|^2, \qquad (7)$$

where $N_\Re$ is the total number of realisations (it's maximum value is equal to the number $N_\xi$ of degrees of freedom, or local modes, of the gravitational protofield, involved in the interaction process [4-17]), $\rho_r(\xi,q) = \left|\Psi_r(\xi,q)\right|^2$ is the generalised density of the *r*-th realisation with the state-function $\Psi_r(\xi,q)$, and the sign $\oplus$ designates the special, dynamically probabilistic meaning of the sum outlined above.

The system state-function $\Psi_r(\xi,q)$ entering the general solution, eq. (7), is obtained in the unreduced EP method in the form [4-17]:

$$\Psi_r(\xi,q) = \sum_i c_i^r \Big[ \varphi_0(q) \psi_{0i}^r(\xi) +$$

$$+ \sum_{n,i'} \frac{\varphi_n(q) \psi_{ni'}^0(\xi) \displaystyle\int_{\Omega_\xi} d\xi' \psi_{ni'}^{0*}(\xi') V_{n0}(\xi') \psi_{0i}^r(\xi')}{\eta_i^r - \eta_{ni'}^0 - \varepsilon_{n0}} \Big], \qquad (8)$$

where $\varepsilon_{n0} = \varepsilon_n - \varepsilon_0$, $\{\psi_{ni}^0(\xi), \eta_i^r\}$ are *r*-th realisation eigen-solutions of the *effective* existence equation (obtained from equation for $\psi_0(\xi)$ in the system (3) by the above Green function substitution):

$$\Big[ h_g(\xi) + V_{\text{eff}}(\xi;\eta) \Big] \psi_0(\xi) = \eta \psi_0(\xi), \qquad (9)$$

the EP operator for the *r*-th realisation is defined by its action,

$$V_{\text{eff}}\left(\xi;\eta_i^r\right)\psi_{0i}^r(\xi) = V_{00}(\xi)\psi_{0i}^r(\xi) +$$

$$+ \sum_{n,i'} \frac{V_{0n}(\xi)\psi_{ni'}^0(\xi)\displaystyle\int_{\Omega_\xi} d\xi'\psi_{ni'}^{0*}(\xi')V_{n0}(\xi')\psi_{0i}^r(\xi')}{\eta_i^r - \eta_{ni'}^0 - \varepsilon_{n0}} \ , \qquad (10)$$

and $\{\psi_{ni}^0(\xi),\eta_{ni}^0\}$ are eigen-solutions of a truncated system of equations:

$$\left[h_g(\xi) + V_{nn}(\xi)\right]\psi_n(\xi) + \sum_{n'\neq n} V_{nn'}(\xi)\psi_{n'}(\xi) = \eta_n\psi_n(\xi). \qquad (11)$$

Note that $n,n' \neq 0$ in eqs. (8)-(11) and everywhere below, contrary to the starting system of equations (3).

The plurality of locally complete solutions of eq. (9), or *dynamic multivaluedness* of the unreduced problem, giving rise to the major property of *causal randomness*, eq. (7), follows from the self-consistent, *dynamically* emerging, *essentially nonlinear*, dependence of the unreduced EP, eq. (10), on the eigen-solutions to be found [4-17]. We thus obtain also the *dynamically derived, a priori probability* $\alpha_r$ of each $r$-th realisation emergence:

$$\alpha_r = \frac{1}{N_{\mathfrak{R}}} \ , \quad \sum_r \alpha_r = 1 \ . \qquad (12a)$$

In the general case, at a higher level of dynamics, we shall have

$$\alpha_r = \frac{N_r}{N_{\mathfrak{R}}} \ \left(N_r = 1,...,N_{\mathfrak{R}}; \sum_r N_r = N_{\mathfrak{R}}\right), \ \sum_r \alpha_r = 1 \ , \qquad (12b)$$

where $N_r$ is the number of "elementary realisations" obtained above and entering the $r$-th actually observed, compound realisation. Note that usual, perturbative models of eqs. (5)-(6) correspond to rejection of all system realisations but a *single*, "averaged" one. We call this property of usual "exact" solutions *dynamic single-valuedness* and the whole standard theory reduction *dynamically single-valued*, or *unitary*, solution and approach.

Another major property of the unreduced solution closely related to dynamic multivaluedness is the *dynamic entanglement* of interacting system components (protofields in this case) expressed by the dynamically weighted products of different component eigenfunctions depending on their respective "degrees of freedom" $(\xi,q)$ in the state-function expression, eq. (8). Dynamic entanglement provides the physical meaning of *interaction* as such, as well as the *rigorous* expression of tangible *quality* of interaction products, absent in any unitary theory describing only an abstract, external "envelope" of a real structure.

The property of dynamic entanglement is further amplified by that of *dynamically probabilistic fractality* of the unreduced solution, which extends essentially the ordinary, dynamically single-valued fractality and is obtained by the repeated use of the same, universal EP method in order to solve truncated systems of equations, starting from eqs. (11), whose solutions enter the expressions for the previous level of structure (see eqs. (8), (10)) [4,8]. One obtains thus the whole hierarchy of not only entangled, but *permanently, probabilistically, interactively changing* and thus *dynamically adapting* realisations of the emerging system structure, which is a *major property* of real structure formation processes, absent in their unitary imitations.

It is not difficult to find the emerging local realisation configuration for two attracting, initially homogeneous protofields [4,5,10,11,13, 17]. The resonant-denominator structure of the state-function and EP expressions, eqs. (8), (10), in combination with the "cutting" integrals in the numerators, shows that the magnitude of the state-function components for each particular ($r$-th) realisation is concentrated around certain eigenvalue $\eta_i^r$ for that realisation, which can be conveniently designated as $\eta_i^r$ and interpreted as the centre of dynamically emerging, local concentration of the attracting protofield density, or *emerging physical space point*. This local *dynamical squeeze* of the initially totally homogeneous protofield system appears to be *inevitable* physically, for the real, *unreduced* interaction dynamics: every small, local density increase of a protofield will provoke a self-amplifying chain of further density increase of both protofields around that location because the larger is the protofield density, the stronger is their attraction at a given place. That omnipresent *dynamic instability* of the unreduced protofield interaction, accompanied and assisted by the above dynamic entanglement, is absent in any unitary approximation, cutting the emerging interaction links and therefore predicting only small deviations from the initial configuration. In the unreduced analysis it leads to maximum local squeeze, or *dynamic reduction*, of the attracting protofields around a location, or (emerging) physical point, which is chosen among other neighbouring, *equally probable* locations in a *causally (dynamically) random* way, in full agreement with the above rigorously derived expressions for realisation structure and probability, eqs. (8), (10), (12). Maximum squeeze of entangled protofields, determining the fully developed structure of a "regular" system realisation, is limited by finite protofield compressibility, and is naturally followed by the reverse process of protofield *disentanglement* and *extension* to the initial, quasi-homogeneous state, which is initiated and governed by the same system instability as the previous phase of reduction.

One obtains thus the emerging, *physically specified* and totally *real* dynamical *structure of (massive) elementary particle*, such as the electron, in the form of *unceasing* periodic cycles of local dynamic reduction and extension of two attracting protofields, where the centre of each next reduction is chosen by the system in a *dynamically (truly) random* fashion among a number of equally probable neighbouring centres. We call this explicitly emerging, internally entangled, highly nonlinear and spatially chaotic particle structure *quantum beat process* [4-6,11-17] (it can also be described as a highly nonlinear and spatially chaotic self-oscillation in the attracting protofield system). Its reality is confirmed by the properties of the unreduced solution within the generalised EP formalism, eqs. (7)-(12). In particular, the latter contains not only the locally squeezed structure of "regular" realisations described above, but also one specific, extended realisation with a "loose" structure (smaller number of contributing eigen-solutions), which describes the disentangled system state during transition between two squeezed, "regular" realisations. It is this transient state called *intermediate, or "main", realisation* that corresponds to effectively weak interaction value of a perturbative approximation (eqs. (5)-(6)) and constitutes the *physically real* particle *wavefunction*, which represents the *totally causal*, physically real extension of the unitary quantum wavefunction (*artificially* mystified because of the *dynamically single-valued* description) and can be further extended, due to the unrestricted universality of our analysis, to *any* kind of system and level of world dynamics (where it takes also the form of generalised *distribution function*) [4-6,12-17]. This physically real, interaction-driven *duality* between squeezed and extended state/phase of the quantum beat process within the elementary particle evokes its another definition as *elementary field-particle* [4,5,11-17].

The emerging *length scale*, $\Delta x$, of the quantum beat process is rigorously defined by the distance between neighbouring regular realisations as given by the eigenvalue separation $\Delta_r \eta_i^r$ for different $r$, found from the unreduced EP formalism, eqs. (9)-(10), $\Delta x = \Delta_r \eta_i^r$. It is the length of the elementary, real *quantum jump* of the squeezed, "corpuscu-lar" state of the particle, or *virtual soliton*, between its consecutive regular realisations within the quantum beat process, actually equal (up to a factor of $2\pi$) to the Compton length $\lambda_C$ for the electron, $\Delta x = \Delta_r \eta_i^r = \lambdabar_C = \lambda_C / 2\pi$ [4-6,11-17]. Another characteristic length scale, determining the *size* of the virtual soliton, or "particle" (electron) as such, is given by the generic eigenvalue separation $\Delta_i \eta_i^r$ for different $i$, equal to the "classical radius" $r_e$ of the electron, $\Delta_i \eta_i^r = r_e$ (see also section 3.2). We thus obtain the *physically real*, naturally *discrete*, *dynamically entangled* and *chaotically* changing *space*.

Since we have obtained the well-defined *events* of dynamic reduction-extension, we can define the *physically real time*, whose *unceasing flow* is derived as *permanent realisation change* of *dynamically multi-valued* protofield interaction process (quantum beat), *intrinsic irreversibility* is provided by the *dynamically random* sequence of realisations (reduction centres), and *elementary interval*, $\Delta t$, can be obtained as $\Delta t = \Delta x / c$, where $\Delta x = \Delta_r \eta_i^r$ is the above space element (elementary jump length) and $c$ is the speed of perturbation propagation in the e/m protofield interacting with the gravitational protofield (known as the *speed of light*). It is clear that $\Delta t = \tau$, where $\tau$ is the period of quantum beat and $\nu = 1/\tau$ is its frequency.

A big number of *different* elementary field-particles will emerge in the described way in the initially homogeneous system of two interacting protofields. This follows from the same basic property of dynamic multivaluedness and its hierarchical fractal structure. Local quantum beat processes can have several major realisations with essentially different EP magnitude, where relatively small amplitudes form the (compound) realisation of lighter particles (leptons) with weaker relation to the gravitational protofield, while much larger amplitudes constitute hadron realisations with closer entanglement of e/m and gravitational protofields. Each of such "big" compound realisations of the first level of interaction structure can contain various particle subspecies and ends up in splitting into numerous individual particles situated at different (emerging) locations and represented by a certain level of the fractal hierarchy of dynamic multivaluedness, described above as the quantum beat process within each (massive) particle.

Higher levels of (weaker) interaction between these entities of the first complexity level start then naturally emerge (see below), but the factor of deep *cosmological* importance at this and higher levels of structure emergence is their *intrinsic, dynamic adaptability* determined by the *self-* *consistent* dependence of the *unreduced*, nonperturbative EP, eqs. (8)-(10), on the emerging structure parameters (exemplified by eigenvalues $\eta$). Thus, any new particle emergence increases the protofield tension, and when the latter is high enough, no more particles can form (for a given interaction magnitude). Therefore the protofield interaction strength dynamically determines the number (mass density) of particles in the universe. One obtains thus a *self-tuning real universe* that avoids, simply due to its unrestricted and realistic interaction problem solution, any "anthropic" problems or "critically adjusted" universal constants [4-6,11-13,17] (see also sections 3.2 and 3.3).

The quantity of *dynamic complexity* as such of *any* real interaction process and emerging structure can now be *universally* defined as a growing function of the total number of its realisations (*explicitly obtained* from the unreduced problem solution) or of their rate of change, equal to zero for the unrealistic case of only one system realisation.[1] It is the latter extreme simplification of reality that is exclusively considered in the unitary theory, including its imitations of "complexity" and cosmology, which explains, as we continue to show below, all its "old" and "new" problems. The physically real, dynamically emerging space and time defined above constitute *two universal, elementary manifestations* of the unreduced complexity, characterising a *single* realisation structure (space) and change/emergence (time). We shall proceed now to major *forms* and *measures* of dynamic complexity, representing *all* system realisations and thus its causally complete structure and dynamics.

A universal measure of complexity is provided by the simplest combination of *independent* space and time elements. It is known as *action* that acquires now an extended, *universal* and *essentially nonlinear* meaning, $\Delta \mathcal{A} = p\Delta x - E\Delta t$, where $\Delta \mathcal{A}$ is the *dynamically determined* action-complexity increment during elementary realisation change, while $E$ and $p$ are coefficients identified as *energy* and *momentum*. They represent universal *differential* measures of complexity related to the *integral* measure of action:

$$E = -\frac{\Delta \mathcal{A}}{\Delta t}\Big|_{x=\text{const}} \quad , \quad p = \frac{\Delta \mathcal{A}}{\Delta x}\Big|_{t=\text{const}} \quad . \tag{13}$$

The action-complexity increment $\Delta \mathcal{A}$ for a field-particle at rest corresponds to one quantum beat cycle and explains the origin of Planck's

---

[1] It is clear that dynamic complexity thus defined is also a measure of dynamical randomness, or chaoticity, or (generalised) entropy (see the end of this section).

constant, $\Delta\mathcal{A} = -h$, after which eq. (13) takes the form:

$$E_0 = -\frac{\Delta\mathcal{A}}{\Delta t} = \frac{h}{\tau_0} = h\nu_0 \ , \tag{14}$$

where $E_0$ is the particle *rest energy*, $\tau_0 = \Delta t$ is the quantum beat period at rest, and $\nu_0 = 1/\tau_0$ is its frequency. Since the rest energy results from spatially *chaotic* wandering of the virtual soliton within the particle wave field, it possesses the *causally substantiated* property of *inertia* as expressed by the *rest mass* $m_0$, $E_0 = m_0 c^2$, where $c^2$ is a coefficient for the moment (rigorously identified later as the square of the light velocity). We can understand now the true meaning of a basic relation used by Louis de Broglie for derivation of his formula for the particle wavelength [23,24] as the expression of chaotic, essentially nonlinear quantum beat dynamics [4,11-17]:

$$m_0 c^2 = h\nu_0 \ .$$

The *state of (global) rest* of a field-particle (or any system) corresponds to the local minimum of complexity-energy and the most homogeneous distribution of realisation probabilities. (Global) *motion* is *rigorously defined* as increased complexity and inhomogeneous realisation probability distribution ($p \neq 0$), so that

$$\frac{\Delta\mathcal{A}}{\Delta t} = \frac{\Delta\mathcal{A}}{\Delta t}\Big|_{x=\text{const}} + \frac{\Delta\mathcal{A}}{\Delta x}\Big|_{t=\text{const}} \frac{\Delta x}{\Delta t} \ ,$$

which transforms eq. (14) into

$$E = -\frac{\Delta\mathcal{A}}{\Delta t} + \frac{\Delta\mathcal{A}}{\lambda}\frac{\Delta x}{\Delta t} = \frac{h}{\text{T}} + \frac{h}{\lambda}\upsilon = h\text{N} + p\upsilon \ , \tag{15}$$

where $E = h/\tau = h\nu$ is the total energy, $\tau = \Delta t|_{x=\text{const}}$ is the quantum beat period of a moving field-particle measured at a fixed space point, $\nu = 1/\tau$, $\lambda = \Delta x|_{t=\text{const}} = \lambda_\text{B} = h/p$ is the space element of the *moving* field-particle, known as *de Broglie wavelength* $\lambda_\text{B}$, $\text{T} = \Delta t$ is the "total" quantum beat period ($\text{T} \neq \tau$), $\text{N} = 1/\text{T}$, and $\upsilon = \Delta x/\Delta t$ is the velocity of global field-particle motion. Since the latter emerges only as the *average tendency* in the *chaotic* virtual soliton wandering with the single jump velocity $c$ (the material *speed of light* defined above), one can express the *thus causally explained* difference between the single jump speed $c$ and the global motion velocity $\upsilon$ by the generalised "relativistic" *dispersion relation* (now rigorously derived) [12,13,17]:

$$p = E\,\frac{\upsilon}{c^2} = m\upsilon\,, \qquad (16)$$

where the *total mass* $m = E/c^2$, now by *rigorously obtained* definition, reflecting the revealed *chaotic* (dynamically multivalued) internal content of *any* (material body) motion and energy. Using eq. (16), one gets the known, but now *causally derived* and realistically explained expression for the de Broglie wavelength:

$$\lambda = \lambda_B = \frac{h}{m\upsilon}\ . \qquad (17)$$

In addition, the dispersion relation thus derived from causal *quantum* dynamics, $p = m\upsilon$, provides (upon time differentiation) the true origin and *rigorous* substantiation of Newton's laws of *classical* mechanics (in their *relativistic* version), thus demonstrating the essential role of underlying *complex* (multivalued) interaction dynamics also at those higher, classical levels of world dynamics.

Using the relation between $p$ and $E$ of eq. (16) and the total energy expression through the quantum beat period ($E = h/\tau$) in eq. (17), we get the *rigorously derived* expression of *time relativity* and its *causal origin* in the underlying complex interaction dynamics:

$$\tau = T\!\left(1 - \frac{\upsilon^2}{c^2}\right)\,. \qquad (18)$$

Time goes more slowly "within" the moving elementary field-particle ($T > \tau$) because the time flow *is produced* by the *same*, *complex-dynamic* (multivalued) *interaction* process that gives rise to the global motion. If we use the straightforward relation to the quantum beat period at rest, $T\tau = (\tau_0)^2$ [4,12,13,17], we get the canonical expression of (now causally derived) time relativity:

$$N = \nu_0\sqrt{1 - \frac{\upsilon^2}{c^2}}\ ,\ \ T = \frac{\tau_0}{\sqrt{1 - \dfrac{\upsilon^2}{c^2}}}\ . \qquad (19)$$

Combination of eqs. (15)-(17), (19) provides now the explicit expression of *unified*, causal understanding of quantum *and* relativistic behaviour of a field-particle obtained as the holistic quantum beat process:

$$E = h\nu_0 \sqrt{1 - \frac{\upsilon^2}{c^2}} + \frac{h}{\lambda_B}\upsilon = m_0 c^2 \sqrt{1 - \frac{\upsilon^2}{c^2}} + \frac{m_0 \upsilon^2}{\sqrt{1 - \frac{\upsilon^2}{c^2}}}, \qquad (20)$$

The quantum wave equations (of Klein-Gordon, Dirac and Schrödinger) can be *derived* from eq. (20) by *causal quantisation*, expressing multivalued dynamics in terms of intermediate, delocalised realisation of the wavefunction [4-6,12,13,16,17] (see also below).

Elementary field-particles, causally obtained thus as complex-dynamical quantum beat processes, form the entities of the first level of emerging real-world structure, or the first *level of complexity*. Due to the *physically* unified world construction of two interacting protofields, the entities of the first level start interacting among them and form higher levels of complex-dynamical world structure by the same, universally described development of unreduced interaction process towards the probabilistic dynamical fractal of the world structure. The number, physical origin, and properties of the four "fundamental forces" between particles obtain a transparent explanation within this theory [5,6,12-17] designated as *quantum field mechanics*. Long-range particle interaction through the e/m and gravitational protofield gives the omnipresent e/m and gravitational interactions, respectively, whereas short-range ("contact") interaction between the protofield elements (poorly resolved as such) appears as "weak" and "strong" interaction forces, where one can clearly see the origin of the (known) unification of e/m and weak interactions (transmitted by the e/m protofield) and similar (but unrecognised) unity between the gravitational and strong interactions. Moreover, *all the four interactions* are naturally, *dynamically unified* within each elementary (hadronic) particle-process, especially in the maximum squeeze state of its unceasing quantum beat pulsation. The physical origin of the gravitational protofield, or medium, can also be causally understood now as a dense, dissipative form of "quark matter" (or "condensate"), where the famous "confinement" of quarks acquires a transparent explanation. Photons, on the other hand, can be interpreted as relatively much weaker, and therefore quasi-regular and massless, excitations of the "elastic" e/m protofield, stabilised by attraction to the gravitational medium (staying thus closer to usual, regular solitons).

One obtains also the dynamic, causal interpretation of *electric charge* (as the *fixed temporal phase* of the quantum beat pulsation), its "quantised" value, and two "opposite" types (as quantum beat *synchroni-* *sation* in the e/m medium) [4,12-17], where the *quantised* e/m interaction by "exchange of photons" (during the "extended" phase of quantum beat) acquires now a *physically real* meaning. The property of *spin* and related *magnetic field* effects are driven by highly nonlinear *vortex* dynamics of the reduction-extension process within every quantum beat cycle.

Further development of complex-dynamical interaction between field-particles leads to causally understood processes of *genuine quantum chaos* (in the absence of dissipation) [4-6,9], *quantum measurement* (small dissipation) [4,10], and *classical (permanently localised) behaviour* emergence in elementary bound, *closed* systems (like atoms) [4-6,12-17], without any extrinsic "decoherence". Classical behaviour emerges as a next, *higher level of complexity* that gives rise, in its turn, to all superior complexity levels by further development of the same unreduced, intrinsically unified interaction process between two initially homogeneous protofields. The *complete macroscopic world structure and dynamics* is thus *explicitly obtained* from that starting "minimal" interaction configuration, where such persisting "cosmological" problems as the origin and emergence of space and time, the "wavefunction of the universe", classicality emergence, and quantum gravity are naturally solved, together with other problems of fundamental physics, within the *intrinsically unified* description of complex interaction dynamics [4-7,12-17].

The unrestricted universality of structure emergence description finds its perfect expression in the *universal symmetry (or conservation) of complexity* [4-6,13,17], which provides the *unified, causally complete* extension of *all* (correct) dynamic equations, laws and principles, remaining otherwise unexplained (postulated), separated, and often contradictory within the dynamically single-valued projection of reality in the standard, unitary theory. The causally specified *qualitative change* and *explicit structure emergence* in the universal science of complexity permit us to introduce two major *forms of dynamic complexity*. One of them is called *dynamic information*, *I*, and expresses the real interaction complexity *before* any structure emergence has actually begun. It generalises the usual notion of "potential energy" and is actually given (in its integral version) by the *generalised action-complexity*, $\mathcal{A}$, introduced above. The second universal form of complexity is called *dynamic entropy*, *S*, and characterises the *unfolded* dynamic complexity of *already appeared*, developed structures (it generalises the usual notions of "kinetic" and "heat" energy).

The *symmetry, or conservation, of complexity* follows from the fact that the system realisation number $N_{\mathfrak{R}}$ determining its complexity $C(N_{\mathfrak{R}})$ is determined itself by the initial mode (combination) number (see above) and thus remains unchanged during the interaction process development. It means that every process occurs so that the sum of dynamic information $I$ and dynamic entropy $S$, or *total dynamic complexity* $C = I + S$, *remains constant*, $C = I + S = \text{const}$, which implies that *always decreasing* dynamic information $I = \mathcal{A}$ (expressing system's "potentialities") is *transformed* into the dual, *always growing* complexity form of dynamic entropy, $\Delta S = -\Delta I > 0$. The "first" and "second" laws of thermodynamics are thus essentially extended to *any* kind of system or process, *unified* in a dynamic *symmetry* and liberated from the skewness of the usual second law (which resolves the related cosmology problems, see section 4). Contrary to any unitary symmetry, the symmetry of complexity is always *exact* (never "broken"), but gives and relates generally *irregular* structures.

The dynamic version of the symmetry of complexity is obtained if we divide its differential expression, $\Delta \mathcal{A} = -\Delta S$ (where $\Delta \mathcal{A} = \Delta I$ and $\Delta S$ are real, finite increments of dynamic information and dynamic entropy), by a (dynamically) discrete time increment $\Delta t|_{x = \text{const}}$, to get the *generalised Hamilton-Jacobi equation* [4-6]:

$$\frac{\Delta \mathcal{A}}{\Delta t}\Big|_{x = \text{const}} + H\left(x, \frac{\Delta \mathcal{A}}{\Delta x}\Big|_{t = \text{const}}, t\right) = 0 \ , \qquad (21)$$

where the Hamiltonian $H = H(x, p, t)$ expresses the entropy-like, differential form of complexity, $H = (\Delta S / \Delta t)|_{x = \text{const}}$, and eq. (13) is taken into account. Because of the *dynamically random* order of emerging realisations, the dynamic information $I = \mathcal{A}$ can only *decrease* with each real time step, which means that the total time derivative of action, or (generalised) *Lagrangian*, $L = \Delta \mathcal{A} / \Delta t = pv - H$, is always negative:

$$L < 0 \ \Rightarrow \ E, H\left(x, \frac{\Delta \mathcal{A}}{\Delta x}\Big|_{t = \text{const}}, t\right) > pv \geq 0. \qquad (22)$$

We obtain in that way the *rigorously derived* expression of the *arrow of time* always oriented, according to eq. (22), in the direction of growing dynamic entropy (and interaction process development). Note that for a system globally at rest ($p = 0$), this condition is equivalent to strict positivity of (generalised) complexity-energy (or Hamiltonian): $E, H > 0$.

The *dynamic, or causal, quantisation* condition describes the unceasing realisation change through the intermediate state of wavefunction, $\Psi$, and means that this state and the total system complexity remain the same after *each cycle* of realisation change [4-6,12,13,15-17]:

$$\Delta\left(\mathcal{A}\Psi\right) \text{ or } \Delta\mathcal{A} = -\mathcal{A}_0\frac{\Delta\Psi}{\Psi} = -i\hbar\frac{\Delta\Psi}{\Psi} \ , \tag{23}$$

where $\mathcal{A}_0$ is a characteristic action value that may contain a numeric constant reflecting interaction details (thus, $\mathcal{A}_0 = i\hbar$, $\hbar = h/2\pi$ at the lowest, "quantum" complexity levels). Combining now eqs. (22) and (23), we obtain the "wavefunctional" counterpart of the universal Hamilton-Jacobi equation in the form of *universal Schrödinger equation* for the generalised wavefunction (or distribution function):

$$\mathcal{A}_0\frac{\Delta\Psi}{\Delta t}|_{x=\text{const}} = \hat{H}\left(x,\frac{\Delta}{\Delta x}|_{t=\text{const}},t\right)\Psi(x,t) \ , \tag{24}$$

where the Hamiltonian operator $\hat{H}$ is obtained from the Hamiltonian $H = H(x,p,t)$ by causal quantisation.

The generalised Schrödinger equation is completed by the *generalised Born rule*, obtained from the *dynamic matching conditions* for regular and intermediate realisations (they give the coefficients $c_i^r$ in the universal state-function expression, eq. (8)) and presenting the wavefunction or its squared modulus as *realisation probability distribution* [4-6,11,13,16,17]:

$$\alpha_r = \left|\Psi\left(x_r\right)\right|^2 \ , \tag{25}$$

where $x_r$ is the $r$-th realisation configuration and one may have the value of the generalised distribution function itself at the right-hand side of eq. (25) for "corpuscular" (rather than "undular") complexity levels. The comparison between eq. (25) and the initial expression for the dynamic realisation probabilities $\alpha_r$, eqs. (12), reveals the universal and *realistic* meaning of the (generalised) wavefunction (desperately missing especially in the unitary quantum mechanics) as the chaotically fluctuating field and state-function of the intermediate system realisation *transformed* into each of its regular realisations (and back), in agreement with eq. (25).

Equations (21)-(25) constitute the basis of the *unified Hamilton-Schrödinger formalism* accompanied by the unreduced, dynamically multivalued equation solution, such as the above result of the generalised EP method, eqs. (7)-(12). This universal formalism is a rigorous expression of the universal symmetry of complexity and unifies extended versions of various particular dynamical equations, usually corresponding to several first terms of power-series expansion of the generalised Hamiltonian [4-6]. It provides also the decisive self-consistent substantiation of the Hamiltonian form of the starting existence equation, eq. (1).

Cosmological meaning of the universal symmetry of complexity goes, however, far beyond its particular mathematical expression. It represents the unified, exact *Order of the World*, applicable to the universe in the whole or any its part, including its *causally specified* origin and structure development (in their *realistic*, unreduced versions). Symmetry of complexity rigorously excludes, in particular, any possibility of universe emergence from "nothing" (with zero total energy), since *only positive* (and big) values of initial interaction complexity (in the form of dynamic information) can give rise to further structure development (with equally positive and big total energy) and real time flow, eq. (22) (see also section 4). This fundamental positivity of the universe content, distinguishing it from the zero-content unitary models, is directly related to the dynamic multivaluedness and intrinsic randomness of any real process, reduced to the dynamically single-valued projection in the unitary schemes that avoid any real, change-bringing interaction. We shall see below that the properties of the unreduced, dynamically multivalued world dynamics permit one to consistently solve, or often do not even contain, the accumulating "new" and stagnating "old" problems of the unitary cosmology and astrophysics, including the "missing" mass and energy content of the world.

## 3. GLOBAL PROPERTIES OF EMERGING COMPLEX-DYNAMICAL UNIVERSE

We shall outline, in this section, the "global", cosmological properties of the real, complex-dynamical world construction, such as they *follow explicitly* from the unreduced, multivalued dynamics of the underlying protofield interaction process (some of them are already mentioned in section 2). Note that practically none of this real-world properties can be consistently reproduced by any version of the unitary theory, irrespective of whether it is recognised as a true cosmological problem or not. Artificial addition of new abstract entities (such as "hidden dimensions" or new, equally "invisible" particle species), accompanied by "suitable" parameter adjustment, certainly cannot change this situation, since new entities create new difficulties, thus simply displacing, or renaming, previous problems that remain basically unsolved because of deceptive reduction to over-simplified, effectively zero-dimensional models. Any observed general, *universal* enough property can be consistently explained only with the help of a *qualitative feature* of the system (interaction) *dynamics* and not by introduction of a new, *specific entity*.

## 3.1. Physically real, 3D space structure and irreversible time flow

We have seen in section 2 how the unreduced interaction between two initially homogeneous protofields gives rise to highly inhomogeneous structure of physically real, *tangible* space and equally real, but *immaterial*, *irreversibly flowing* time that can *not* be *really* "mixed" with space in an (abstract) "manifold".

We causally derive the *exact number (three) of spatial dimensions*, or "degrees of freedom", as being due to the dynamic entanglement of two protofields and their physically real interaction as such. This conservation of the number of basic entities, or "degrees of freedom" during the interaction process is the meaning of the universal symmetry (conservation) of complexity (see the end of section 2) supported by the *totality* of existing observations. We thus reveal also the detailed *physical nature* of those *emerging* space "dimensions" (remaining only abstract symbols in the canonical theory): they are obtained as *interaction-driven, chaotically changing, dynamically discrete and fractal entanglement*, or "mixture", of the physically real, initially homogeneous protofields.

We reveal the role of *essential nonlinearity*, omnipresent *dynamic instability* and resulting *causal randomness (chaoticity)* of *quantum beat* dynamics of interacting protofields in establishment of *spatially chaotic* sequence of reduction-extension *events* within each field-particle, which gives rise to *unceasing* and *objectively unpredictable* in detail (and *therefore irreversible*) *time flow*.

*Universality* of the obtained concept of space and time is supported by its unrestricted applicability to any system or level of complexity, giving rise to the fractally structured *hierarchy of space and time* reproducing the hierarchy of world (interaction) complexity and demonstrating the dynamic origin and connection between space and time elements at each level. All cosmological problems of time (its absence in the effectively empty world, magically "tunneling" from nothing, etc.) are thus consistently solved (see the "time flow" condition of eq. (22) and section 4.1 for more details). Another aspect of time and space universality refers to similarity of their fundamental properties throughout the whole "physically infinite" universe. Silently postulated in the canonical theory, this very *special* property finds now its substantiation in the *physically unified* structure of the underlying protofield system and related complex-dynamic *synchronisation of all individual quantum beat processes* (up to phase inversion), which determine the real time flow [4,12,13,17].

## 3.2. Unified complex-dynamical origin of particles, interactions and constants

It is important that the two omnipresent, "pervasive" *manifestations* of un-reduced dynamic complexity, space and time, emerge in the protofield interaction process *in intrinsic unity* with the simplest *structures* of the first level of complexity, *elementary field-particles*, and their *fundamental properties* (mass, energy, motion, electric charge, spin, etc.), particle *interactions* with their observed properties (number, range, relative magnitude, unification), and *dynamical laws* (quantum and classical mechanics, special and general relativity), *all* of them being now *causally* and *explicitly obtained (derived)* from the fundamental interaction dynamics (*without any* "postulates") and thus naturally *unified* (section 2) [4,5,11-17]. The fundamental (measured) properties of real world structures are related *measures* of the *same*, *universally* defined *dynamic complexity*, while structures themselves and their interactions represent two universal, dual *forms of complexity*, dynamic entropy and dynamic information, respectively, which are *permanently transformed* into one another according to the underlying *unique* "order of the world", the universal *symmetry of complexity*. Omitting here the detailed discussion of this *intrinsically unified* world structure and dynamics (section 2), we note only the *indispensable* role of omnipresent *dynamic multivaluedness* and the ensuing *chaoticity*, *diversity* (multiplicity) of forms and *adaptability* of real interaction products (absent in *any* unitary model), starting from the *quantum beat* process that constitutes the *causally complete* structure of (massive) elementary particles.

The related "difficult" problems of the unitary cosmology, which are *naturally solved* in our complex-dynamical description, include the *problem of the universe wavefunction*, *quantisation of gravity*, and quantum cosmology. The universe wavefunction is *causally specified* now as the intermediate realisation of quantum beat processes in the *physically unified* protofield system. It naturally loses its *quantum* meaning there where classical (bound) systems start to emerge, but the *generalised* wavefunction and Schrödinger equation (see the end of section 2) re-emerge at each higher complexity level. As for the problem of *quantum* gravity, our universal gravitation is an indirect relation between *naturally discrete* quantum beat processes through the gravitational protofield and has therefore *causal (complex-dynamic) quantum origin* from the beginning (as well as *the entire universe*) [4,5,12,13,17].

An essential novelty of the complex-dynamic cosmology is that it shows the *physical* origin of *universal constants* and their *universality*, reduced eventually to the physically unified origin of the universe.

We have seen above (section 2) that one of the constants, the *speed of light c*, is introduced in our theory not as an abstract, postulated "limit to signal speed" (standard relativity), but as a "normal", physical speed of signal propagation in the e/m protofield coupled to the gravitational medium, while time relativity and related limit to signal propagation velocity are consistently *derived* from the underlying *complex* interaction dynamics [4,5,12-17].

The (new) physical origin of the *fine structure constant* α follows from a new form of the well-known relation between $\hbar = h/2\pi$, α, and elementary charge $e$, involving the electron rest mass $m_0$ and the Compton wavelength $\lambda_C = h/m_0 c$ :

$$\alpha h = \frac{e^2}{c} \ \Rightarrow \ m_0 c^2 = \frac{2\pi}{\alpha} \frac{e^2}{\lambda_C} = N_\Re^e \frac{e^2}{\hbar_C}, \ \ \hbar_C = \frac{\lambda_C}{2\pi}, \qquad (26)$$

where $N_\Re^e = 1/\alpha \approx 137$ emerges now as the *realisation number of the electron* as a complex-dynamic interaction process (quantum beat), so that the fine structure constant α coincides with its realisation probability $\alpha_r$ (see eq. (12a)), $\alpha = \alpha_r$, while $\hbar_C = \lambda_C/2\pi$ is the "quantum jump" length of the virtual soliton. We can rewrite eq. (26) also as $\hbar_C = N_\Re^e r_e$, where $r_e = e^2/m_e c^2$ is the usual "classical radius" of the electron, which means that the *size of the virtual soliton $D_e$* can be estimated as $D_e = 2\pi r_e = \pi d_e$, $d_e = 2r_e$ being the classical electron diameter/size.

The true physical origin of *Planck's constant $\hbar$* follows from another form of eq. (26) and the Compton wavelength expression:

$$\hbar = \hbar_C p_0 = N_\Re^e \frac{e^2}{c}, \qquad (27)$$

where $p_0 = m_0 c = E_0/c$. We see that Planck's constant $\hbar$ measures, in units of action-complexity, the "volume" of the protofield EP well, eventually for *any* field-particle, with the width of $\hbar_C$ (or $N_\Re$) and the depth of $p_0$ (or $e^2/c$). This result explains the causal origin of $\hbar$ *universality*, remaining totally "mysterious" in the standard theory, as another manifestation of the universal symmetry of complexity: the protofield deformation for various particles and (united) processes occurs so that the EP well

"volume" in terms of action-complexity, $\hbar$, remains the same (for the fixed protofield interaction and material), whereas its depth (particle mass or charge) and width (realisation number or elementary wavelength) can vary considerably. This rule is additionally confirmed by the related causal explanation of the *largest (quasi-stable) nuclear mass* as being roughly equal to that of the heaviest elementary particle ($\sim 100$ GeV) [15,17].

Finally, the *universal gravitational constant* $\gamma$ of classical Newton's law of gravitation is used, together with $\hbar$ and $c$, in the canonical expressions for Planckian units, underlying many basic constructions of the scholar cosmology and particle theory and giving hugely exaggerated, too big or small, fundamental units of length, time, and mass, separated by many orders of magnitude from the observed extreme particle properties (the "hierarchy problem"). We can see now the origin of those contradictions and genuine involvement and meaning of gravity constant: whereas Planckian units describe *individual* EP well (quantum beat) dynamics *within each particle*, the usual gravitational constant expresses the *indirect* and therefore much weaker interaction between *different* particles (quantum beat processes) by transmission through gravitational medium (hence the famous exceptional "weakness" of the gravity force, always badly understood in conventional theory). Therefore one should use another, effective value of "gravitational constant", $\gamma_0$, in the Planckian unit definition, expressing the magnitude of the *direct*, much stronger *protofield attraction* as the *dynamically unified* origin of *all* interactions, realised in the squeezed state of virtual soliton. It gives just the right values for Planckian units of length $L_P$, time $T_P$, and mass $M_P$, equal to the *observed* extreme values of real particle properties $l_{exp}$, $t_{exp}$, and $m_{exp}$:

$$L_P = \sqrt{\frac{\gamma_0 \hbar}{c^3}} \simeq 10^{-17} - 10^{-16} \text{ cm} = l_{exp} \ ,$$

$$T_P = \sqrt{\frac{\gamma_0 \hbar}{c^5}} \simeq 10^{-27} - 10^{-26} \text{ s} = t_{exp} \ , \qquad (28)$$

$$M_P = \sqrt{\frac{\hbar c}{\gamma_0}} \simeq 10^{-22} - 10^{-21} \text{ g} (10^2 \text{-} 10^3 \text{ GeV}) = m_{exp} \ ,$$

where the relation between $\gamma_0$ and $\gamma$ can be specified, for example, using the values of ordinary Planckian unit of length $l_P$ and measured length $l_{exp}$: $\gamma_0 = (l_{exp}/l_P)^2 \gamma \simeq (10^{33} - 10^{34}) \gamma$.

The "hierarchy problem" is resolved thus without any additional, abstract and unobservable entities (e.g. "hidden dimensions" in "brane-world" models [18-20]), which inevitably create new difficulties and actually replace *dynamic* dimensions of the *multivalued* reality, incorrectly reduced in its single-valued imitations. One can easily deduce from here *major (fatal) consequences* for the parts of standard theory relying upon (usual) Planckian units, such as cosmological *inflation* and *quantum gravity* theories, as well as obtain the causal explanation of the relative *weakness of gravity* (as being due to the small ratio $\gamma/\gamma_0$ ), *dynamic unification of all fundamental forces*, and *causal theory of "black holes"* and other dense "quantum condensates" (section 3.3) [4].

## 3.3. Self-tuning universe structure formation by unreduced interaction adaptability

It is evident already in terms of general logic that a *dynamically emerging* universe should have a dynamically consistent, *self-tuning*, adaptable structure, since this is the *essence* of genuine, *autonomous* structure formation as such. No wonder that this is the property of complex-dynamic universe structure *explicitly* obtained as a result of protofield interaction process (section 2), as it is demonstrated by the *dynamic* origin of major entities, properties and universal constants (section 3.2). Moreover, this *universal* property of the unreduced complex dynamics is preserved at *any higher level* of the emerging world structure. By contrast, it is *impossible* to obtain a feasible, stable universe structure in any unitary model, since its effectively zero-dimensional space leaves no possibility for intrinsic adaptability. Mechanistic adjustment of artificially introduced entities and parameters can provide only a basically inefficient substitute for dynamical tuning, giving the well-known "anthropic" difficulties.

As can be seen from the self-consistent structure of the unreduced EP formalism (eqs. (7)-(12)), a viable universe with the same basic properties will always emerge for generic protofield interaction parameters. According to the universal symmetry of complexity (section 2), greater quantities of dynamic information (generalised "potential energy") in the initial system configuration $V_{init}$ will lead to greater dynamic entropy (generalised mass-energy) of the emerging universe structure, $M_{univ}$ :

$$V_{init} = M_{univ} c^2 \ ,$$

where the emerging structure quickly ramifies into probabilistic (multi-valued) fractal hierarchy of higher complexity levels, maintaining the same principle of intrinsic adaptability:

$$M_{\text{univ}} \rightarrow \sum_{\text{part}} N_{\text{part}} m_{\text{part}} + \frac{V_{\text{fund}}}{c^2} \rightarrow \sum_{\text{atom}} N_{\text{atom}} m_{\text{atom}} + \frac{V_{\text{chem}}}{c^2} \rightarrow \dots , \quad (29)$$

with "part" and "atom" designating progressively emerging species of elementary particles (together with their interaction complexity $V_{\text{fund}}$), atoms (and their interaction complexity $V_{\text{chem}}$), and so on. Since both $V_{\text{fund}}$ and particle masses at the first complexity level depend (through the protofield tension) on the number of particles formed, the latter will be limited quantitatively and qualitatively (in the number of stable particle species). While quantitative aspect is more evident and corresponds to a general balance of eqs. (29), qualitative aspect provides the *causal* explanation of observed *instability* of *all particle species but a couple* of one shallow-EP (leptonic) species, known as the electron, and one deep-EP (hadronic) species, represented by proton.

Exceptions from generic results can exist rather for extreme values of protofield interaction magnitude, but they also find their suitable places in the holistic complex-dynamical world picture.

Ultimately strong protofield interaction will create a macroscopically large, "many-particle" protofield "collapse" that may have a number of different phases [4], from a partially coherent "condensate" of elementary particles ("superdense" cosmic objects, such as "neuron stars"), which is still a part of "ordinary" reality, to the total protofield collapse down to their "pre-interaction" state of the unique "proto-matter", which does not contain anything from this world and should be considered as effective nothingness with respect to its structure. Contrary to abstract and contradictory, finally postulated "exact solutions" of the unitary theory (such as "black holes"), each of these states can be provided with the causal, physically specified origin and structure, showing qualitative correlations with a number of observed "exotic" objects of the universe (e.g. quasars) and their specific features.

The case of ultimately weak protofield interaction corresponds to small fluctuations of their structure that cannot transform to real, massive matter and may account for either "primordial" state of the protofields or, more realistically, the observed universe state away from massive matter, in the "vacuum", including propagating ordinary photons and, in particular, the "microwave radiation background" related in the standard cosmol-ogy to the "remnants" of the first stages of the Big Bang.[2] We see now that in the causally emerging, interaction-driven universe structure such "vacuum fluctuations" (cf. also "zero-point field") are inevitable and need not be related to a specific cosmological "scenario" or imposed abstract entity (see also [25]).

Note finally the huge, *exponentially large efficiency* of complex-dynamic adaptability (self-tuning) process: it is due to unceasingly breeding and permanently changing realisations of the probabilistic dynamical fractal (section 2), which gives rise to real-time, "fantastically efficient" exploration by the system of (almost) all existing possibilities for structure development [5].

## 4. UNIFIED SOLUTION TO THE PROBLEMS OF MASS, ENERGY AND ENTROPY

### 4.1. Universe energy positivity and the dynamic time arrow

According to the universal symmetry (conservation) of complexity (see the end of section 2), the total dynamic complexity does not change in a structure emergence process, but is transformed instead from its "latent" (but real and *positively* defined) form of dynamic information (expressed by the generalised action) into the "unfolded" form of dynamic entropy. Therefore any "compensation" of positive total energy of moving bodies by negative energy of their gravitational attraction, as it is implied by the unitary cosmology, is impossible in the real world dynamics. In fact, this "zero-energy balance" is due to zero-complexity reduction of the dynamically single-valued model of the standard theory. By contrast, the inevitable positivity of the total complexity-energy of any real system is due to its dynamically multivalued, and therefore chaotic, dynamics, where the "thermal energy" of chaotic realisation change always determines the large positive balance of the total energy.

This energy positivity condition is directly related to the direction of the arrow of time (and the very existence of time flow), by a rigorously derived and absolutely universal relation of eq. (22), which means that the *positive* stock of total energy-complexity gives rise to the *flow of time* as such, since for the system globally at rest $\Delta t = -\Delta \mathcal{A}/E$, and with $\Delta \mathcal{A} < 0$ (because of dynamic multivaluedness) $\Delta t > 0$ only if $E > 0$. In other words, a universe with zero total energy *could not exist* at all, in any configuration. Moreover, a small positive energy will give rise to proportionally small mass-energy content of the universe (see also section 3.3). This fundamentally substantiated conclusion about the *real, dynamically multivalued* universe emergence and structure puts an end to various formal postulates and hypotheses of unitary cosmology about possibility of universe appearance from nothing by a sort of "quantum tunneling" or "vacuum fluctuation", based on the zero energy balance (where positive mass-energy of "matter" is compensated by negative energy of gravitational attraction). It involves also the popular "Hamiltonian constraint", applied e.g. in the unitary "quantum cosmology" (including the Wheeler-DeWitt equation). Even when unitary theory inserts a positive energy in its formally postulated equations, it does not see the genuine physical origin and meaning of both energy/mass and its positivity, losing the main, chaotic part of system dynamics. Indeed, the zero energy balance is impossible because the dynamically multivalued, chaotic part of any dynamics adds the dominating positive part to the total energy. We shall see that this loss of the main part of energy and motion in the unitary theory underlies *all* "difficult" problems of cosmology and astrophysics: mass and energy are lost in the unitary universe models *from the beginning*, and there is no wonder that various aspects of this basic deficiency emerge inevitably with the growing precision and completeness of measurements.

---

[2] Note that unitary theory often makes reference to "vacuum fluctuations" of "zero-point field" or "space-time foam" obtained as *formal* solutions of eventually *postulated* equations. We emphasize the causal origin of our weak interaction limit within the *same*, unique interaction process between two protofields at small values of *effective* coupling, where any strong protofield deformation, and therefore quantum beat dynamics, is impossible.

Another aspect of positive complexity-energy and time arrow of a real universe is the permanent, *strictly positive growth of dynamic entropy* accompanying *any* structure emergence, which resolves the old contradiction of unitary theory between the "second law" (entropy/disorder growth) and apparently "growing order" during structure formation. Any unitary structure is basically regular only because of artificial limitation (dynamic single-valuedness) of the unitary theory itself, while the unreduced analysis of structure creation process shows (section 2) that any, even most externally regular structure, can appear and exist only due to the dominating *internal chaoticity* of its different, though maybe quite similar realisation change (which is a limiting regime of "multivalued self-organisation") [4-7]. It is yet more important that this omnipresent entropy growth constitutes only a part of the *symmetry*, or conservation, of complexity (again contrary to the unitary science paradigm), since it occurs at the expense of equal decrease of the initial dynamic information of the system interaction configuration. The universe, its *real* structure, evolution, and any part dynamics are based therefore on the absolutely general and *exact* (never broken) principle of symmetry, the symmetry of (unreduced) complexity, constituting thus the genuine *Order of the World* that possesses the intrinsic, autonomous, rigorously specified *structure creation power*.

## 4.2. Locally missing mass: Unitary model deficiency

The so-called *dark mass problem* involves various observation data showing that *local* cosmic structure dynamics (mostly for galaxies) would need much larger (from several to hundreds times more) quantities of ordinary, massive matter, than those that can actually be perceived (see e.g. refs. [26-29]). Big variability of the missing mass effect is an equally puzzling feature of the problem. We show that these difficulties of the unitary theory originate from the same incorrect neglect of the main, chaotic part of system dynamics, now occurring at the level of local cosmic object interaction. If one considers the real, dynamically multivalued system behaviour, the problem will not appear and the truly chaotic dynamics of real objects will account for the observed dynamical features with the "visible", normal mass values. It is important that one should take into account the *genuine, dynamically multivalued* chaos, rather than the one of unitary imitations by "involved" but basically regular (and unique) trajectory.

The main idea is physically straightforward: because of artificial cut of all system realisations but one in the unitary theory (this is an *exponentially big* reduction for a many-body system), one obtains inevitably a "missing motion" problem, which is *interpreted* as a mysteriously "missing mass" within the same unitary imitation. One can specify this conclusion in various ways, and we start with a demonstration of incompleteness of the standard "virial theorem" application to the real, multivalued dynamics of a many-body system, since it shows how the major "balance" between potential and kinetic energy can be modified by the *true* chaos.

If system components move under the influence of gravitational attraction, e.g. in a galaxy, then the ordinary virial theorem gives the following relation between the time-averaged values of kinetic $\overline{T}$ and potential $\overline{U}$ energy of a system or any its component (see e.g. [29]):

$$2\overline{T} = -\overline{U}, \qquad (30)$$

whereas in reality this *regular* kinetic energy, $\overline{T} = \overline{T}_{\text{reg}}$, is a *small* part of its true, *chaotic* content $\overline{T}_{\text{real}}$:

$$\overline{T}_{\text{real}} = \overline{T}_{\text{reg}} N_{\mathfrak{R}}, \tag{31}$$

where $N_{\mathfrak{R}}$ is the *effective* number of system realisations for a given type of observation and respective "averaging" (usually $N_{\mathfrak{R}} \gg 1$, while $N_{\mathfrak{R}} = 1$ for the unitary model of the standard theory).

The *observed* potential energy, $\overline{U}_{\text{obs}}$, gives *real* kinetic energy:

$$2\overline{T}_{\text{real}} = -\overline{U}_{\text{obs}}, \tag{32}$$

but when observations are interpreted within a unitary, deficient version of dynamics, eq. (30), stating that

$$2\overline{T}_{\text{reg}} = -\overline{U}_{\text{obs}}, \tag{33}$$

one obtains a discrepancy, $\delta$, dividing eq. (32) by eq. (33):

$$\delta = \frac{\overline{T}_{\text{real}}}{\overline{T}_{\text{reg}}} = N_{\mathfrak{R}}. \tag{34}$$

It is explained *within* the unitary model as being due to the "invisible", but actually present, or "dark", mass, $M_{\text{dark}} = M_{\text{real}} - M_{\text{reg}}$, whose relative value can be estimated as

$$\frac{M_{\text{real}}}{M_{\text{reg}}} = \frac{\overline{T}_{\text{real}}}{\overline{T}_{\text{reg}}} = \delta = N_{\mathfrak{R}}. \tag{35}$$

The observed discrepancy can actually be used, within the unreduced, *complex-dynamic interpretation*, for estimation of effective $N_{\mathfrak{R}}$ values. Since $\overline{T} \propto M\overline{v^2}$, one can say that in reality there is *too much motion*, or (deviating) *velocity*, in a system with respect to unitary expectations, so that one has rather a "dark velocity" effect:

$$\left(\overline{v^2}\right)_{\text{real}} = N_{\mathfrak{R}} \left(\overline{v^2}\right)_{\text{reg}}. \tag{36}$$

One can easily refine this result for a distance-dependent case, $N_{\mathfrak{R}} = N_{\mathfrak{R}}(r)$ (where $r$ is a coordinate within the system), in terms of velocity-distance dependence curves, or "rotation curves", for galaxies. In that case an "anomalous" $v(r)$ dependence is not due to anomalies of mass distribution, $M(r)$ (attributed to "dark matter halos"), but due to "unexpected" (in the unitary model) contribution to *average* velocity from chaotic motion parts, so that in reality $\upsilon(r)$ is proportional not to $\sqrt{M_{\text{reg}}(r) + M_{\text{dark}}(r)}$, but to $\sqrt{N_{\mathfrak{R}}(r)}$. In a general case,

$$\upsilon(r) = \sqrt{\frac{\gamma N_{\mathfrak{R}}(r) M_{\text{obs}}(r)}{r}} \quad \text{or} \quad N_{\mathfrak{R}}(r) = \frac{r\upsilon^2(r)}{\gamma M_{\text{obs}}(r)} \ , \tag{37}$$

where $M_{\text{obs}}(r) = M_{\text{real}}(r)$ is the *ordinary*, "visible" mass within radius $r$, and one can *derive* the features of *chaotic* system dynamics, $N_{\mathfrak{R}}(r)$, from the observed $\upsilon(r)$ and $M_{\text{obs}}(r)$ dependences for perceivable, "normal" object components.

As should be expected, $N_{\mathfrak{R}}(r)$, and thus chaoticity, will typically have a wide, often irregular maximum in "looser" system parts, such as galactic halos or central, inter-component regions of a cluster. It correlates also with the *empirically based* MOND *postulate* that tends to interpret "unusual" motion in those weaker interaction regions as (unexplained) fundamental modification of Newtonian gravitational attraction or dynamics (see e.g. [30]). There is even a deeper link here with our unreduced EP approach: in a real many-body system one always deals with an *effective*, rather than direct, interaction that bears the self-consistent influence of *all* system components, *differs* essentially from the direct interaction and possesses *many* contributing, chaotically changing realisations.

The observed big variations of "dark mass" effects for different objects represent a "heavy" difficulty for any explanation in terms of additional, "invisible" entities, but are, on the contrary, *inevitable* for the above *unified* explanation in terms of the true (multivalued) chaos effects that *should* vary a lot. Moreover, one can trace a definite qualitative correlation between the expected object chaoticity (degree of irregularity), its spatial dependence, and the observed magnitude of "missing mass" effects (extended verification is certainly necessary). One may note also that it is much more consistent to explain an observed, *variable* system *property* by a fundamental *property* of its *dynamics*, rather than by a new, strangely escaping and inevitably *fixed entity* (it refers also to related interpretation of the *origin of mass* in the universal science of complexity and unitary field theory [4,5,12-17]). One should also take into account the spatial dependence of chaotic mass distribution effects (or "structural" chaos) that tend to accumulate just outside of the main mass and interaction concentration in the system (especially the one with a "centred rotation" configuration), in agreement with data interpretation using eqs. (37).

Note finally the discovered *conceptual relation* between the missing mass effects at different levels of world dynamics, including the missing (total) mass-energy of the universe (section 4.1), missing *dynamic* origin of particle mass (replaced by the artificially introduced *new entity* of "Higgs boson"), and "dark mass" effects at the level of cosmic objects, *all* of them explained in the universal science of complexity by the *unified*, rigorously derived, *complete* solution of the *unreduced* interaction problem (cf. section 1).

## 4.3. Globally missing energy and Big Bang contradictions: Deficient linearity

The origin of *globally* missing, "distributed" energy, or "dark energy" [26-28], that could also be called "missing universe acceleration", is directly related to the vicious circle of the unitary cosmology scheme centred on the Big Bang hypothesis or "exploding vacuum" solution. Indeed, the latter starts from *postulated*, artificially imposed *nothingness of the essential universe content* (section 4.1), in the form of dynamically single-valued, zero-complexity reduction of universe dynamics (irrespective of particular "model" details). Because of the intrinsic *instability* of that fundamentally fixed, static construction, one is obliged to further impose a mechanistic "general expansion" (or the reverse squeeze) of the universe as a single possible mode of its (totally illusive) "development". The choice for expansion, or Big Bang, is justified by a *particular interpretation* of the observed "red shift" effect (the interpretation that involves a number of serious contradictions in itself). However, the *conceptual instability* of *any* unitary model (absence of evolving, adaptable *degrees of freedom*, as opposed to abstract "parameters") persists in the form of multiple particular problems of the Big Bang model whose proposed "solutions" only transfer the difficulties to other formulations or artificially introduced entities. The "dark energy" problem represents only the latest in the list, though scandalously big and long hidden rupture in the basically frustrated construction: a *slightly* uneven red-shift dependence on distance leads to a *huge deficiency* in the source of uneven expansion, supposed to be a distributed stock of mysterious, *invisible* energy that should inevitably take *very exotic*, normally *impossible* forms. This final impasse of missing energy (and mass) content of the universe simply takes us back to the beginning of the unitary vicious circle, where such emptiness of the universe content has been *explicitly imposed* by the unitary paradigm itself (this is but another, degenerate case of complexity conservation law, astonishing in its long-lasting simplification, $0 = 0$, applied here to the *whole universe content*).

By contrast, the unreduced, dynamically multivalued and probabilistically fractal structure of real interaction dynamics leads to the *globally stable* concept of universe structure development, just because it is based on the *omnipresent* and massively adaptable *local*, dynamic instability of *explicit structure creation*. The universe structure emergence in the initially homogeneous system of interacting protofields, starting from the physically real space, time and elementary particles, intrinsically unified with their fundamental properties and interactions, can be described as a distributed *im*plosion of ubiquitous, fractally structured creation, as opposed to mechanistic and intrinsically *destructive ex*plosion of the unitary Big Bang (and "inflation") schemes.

Therefore the "dark energy" problem *does not even appear* in the complex-dynamic, intrinsically creative cosmology. The self-tuning universe structure, liberated from unitary instabilities and related "anthropic" speculations, emerges naturally and self-consistently, simply due to the unreduced, truly "exact" picture of the underlying interaction processes. As to the origin of the observed red shift effect in *radiation* spectra of *distant* objects, it finds its consistent explanation in terms of *nonlinear radiation propagation* properties in the system of coupled protofields, where some (relatively weak) loss of energy by soliton-like photons propagating in the e/m protofield medium is *inevitable* because of their *irreducible*, though relatively weak, coupling to the gravitational medium.

Note the essential difference of this *nonlinear* energy dissipation from linear scattering effects in any ordinary, "corpuscular" model. The soliton-like photon, remaining *stabilised* by interaction with the gravitational protofield, can slowly give its energy to the gravitational degrees of freedom *without* any noticeable change of its direction of propagation (i.e. without any "blur" effects in the distant object images). Characteristic "transpiercing" and "circumventing" modes of soliton interaction with "weak" enough obstacles can explain anomalously small loss and vanishing angular deviation effects for photons and very high-energy particles (see below). One should also take into account possible contribution from modified protofield parameters around big mass concentration or various "singular" objects, as well as at earlier stages of universe structure development. Detailed calculations of the effect will inevitably involve many unknown parameters of the system, but qualitative properties and consistency of the whole picture provide convincing evidence in favour of this kind of *fundamentally new* explanation of the red shift effect (within a broader scope of "tired light" approach) and its expected refinement, including the necessary clarification of the detailed *physical origin of photon* (missing persistently in the unitary theory framework).

In particular, the *nonlinear* red shift dependence on distance that gives rise to *catastrophic* consequences in the unitary cosmology can only be *natural* in the complex-dynamical, *essentially nonlinear* picture (section 2). The nonlinear energy-loss mechanism of soliton-like photons explains why this loss grows *more slowly* with distance, than any usual mechanism of diffuse scattering would imply (cf. the above note on soliton scattering dynamics). Similar dynamics could solve, by the way, the persisting puzzle of GZK effect for the ultra-relativistic particles, since at those super-high energies the motion of a massive particle approaches that of (a group of) photons, according to the results of quantum field mechanics [4,12-17]. Another, though maybe less specific feature of red-shift data correlating with our explanation is the apparent *growth of average scatter* of data points with distance.

## 5. CONCLUSION

Returning to the general picture of our emerging universe, note once more that it does not contain "motion-on-circles" dynamics, on any scale of structure creation, so that the initial amount of dynamic information, in the form of protofield interaction, gives rise to the generalised, complex-dynamical system *birth*, followed by its gradual, *irreversible* and "global" transformation into dynamic entropy (developed structure) representing a universally defined, *finite* system *life*, which ends up in the state of generalised *death*, or *equilibrium*, around the total transformation of the initial dynamic information into entropy (unless additional dynamic information is introduced into the system) [4].

The generalised "potential energy" of interacting protofields can be introduced e.g. by their explicit separation from the "pre-existing" state of "totally unified" (mixed) protofields that could have the form of a generally inert quark-gluon "condensate" in its "absolute" ground state (but these "prehistoric" assumptions are subject to inevitable uncertainty and can be estimated rather by general consistency and parsimony principles). What appears to be much more certain, however, is that one *does* need an initial form of "potential" interaction complexity, *positively* defined and speci-fied here as "dynamic information", since the birth of a structured, real universe from absolute "nothingness", without *genuine* interaction development (which is the preferred dogma of conventional unitarity), contradicts the fundamentally substantiated and *universally* confirmed symmetry (conservation) of complexity.

Finally, we may summarise other empirical perspectives of our complex-dynamical universe description, whose consistent development within the standard, unitary cosmology paradigm seems much less probable. The highly uneven, long-distance concentration of various anomalous, super-intense sources of energy (as well as their "peculiar" red-shift tendency) points to a (probably moving) "shape of the world", which looks quite natural in our interacting protofield logic, while it would need additional, "unnatural" assumptions in the Big Bang logic of "exploding emptiness". Growing problems with the universe age can be naturally solved in our complex-dynamic cosmology as it traces explicitly the real life-cycle events of emerging structures, while the unitary theory encounters here another series of its inbred "instabilities" (due to the rigidly fixed "models" and mechanistic data fit). The same refers to structural difficulties of the omnipresent expansion and natural elimination in our approach of this and other "old" difficulties of the unitary theory, such as average space flatness and homogeneity (see also sections 3 and 4). Intrinsic inclusion of realistic, unified solution of the stagnating problems of quantum mechanics, field theory and relativity (sections 2 and 3) into cosmology constitutes the *unique* feature of our theory that, being indispensable, cannot be even expected for any unitary model. Irreducibly complex dynamics of detailed formation and evolution of galaxies, stars and planetary systems is one of the main further applications of the present theory.